\documentclass[
  paper       = a4,
  headheight  = 16pt,
  footheight  = 16pt,
  fontsize    = 10pt,
  twoside     = true,
  titlepage   = true,
]{scrartcl}

\usepackage[utf8]{inputenc}
\usepackage{XCharter}
\usepackage{overpic}
\usepackage{comment}
\usepackage{physics}
\usepackage{enumitem}
\usepackage{booktabs}
\usepackage{capt-of}

\usepackage[ngerman,british]{babel}
\usepackage[english=british]{csquotes}
\setquotestyle[guillemets]{german}

\usepackage[
  backend=bibtex8,
  natbib=true,
  citestyle=authoryear,
  style=authoryear,
  maxcitenames=1,
  maxbibnames=5,
  uniquename=init,
  giveninits=true,
  sortlocale={de_DE_phonebook},
]{biblatex}

\DefineBibliographyExtras{british}{%
}

\DeclareFieldFormat{eprint:urn}{%
  \textsc{\scriptsize{URN\addcolon}}\space
  \ifhyperref
    {\href{http://www.nbn-resolving.org/#1}{\nolinkurl{#1}}}
    {\nolinkurl{#1}}
}

\DeclareFieldFormat{eprint:hdl}{%
  \textsc{\scriptsize{HDL\addcolon}}\space
  \ifhyperref
    {\href{http://hdl.handle.net/#1}{\nolinkurl{#1}}}
    {\nolinkurl{#1}}
}

\bibsetup{
  \setcounter{abbrvpenalty}{10000}
  \setcounter{highnamepenalty}{10000}
  \setcounter{lownamepenalty}{10000}
}

\AtEveryBibitem{\clearfield{month}}
\AtEveryBibitem{\clearfield{day}}

\usepackage[%
  paperwidth  = 17cm,
  paperheight = 24cm,
  top         = 25mm,
  bottom      = 24mm,
  outer       = 15mm,
  inner       = 23mm,
  footskip    = 10mm,
]{geometry}
\usepackage{xcolor}

\usepackage{setspace}
\usepackage{graphicx}
\usepackage{subfig}

\usepackage{placeins}

\usepackage{mathtools}
\usepackage{amsfonts}
\usepackage{amssymb}

\usepackage[font={footnotesize}]{caption}
\setcapwidth{0.8\textwidth}
\setcapindent{0pt}
\addtokomafont{caption}{\centering}
\addtokomafont{captionlabel}{\bfseries}

\colorlet{commentcolour}{green!50!black}
\colorlet{stringcolour}{red!60!black}
\colorlet{keywordcolour}{magenta!90!black}
\colorlet{exceptioncolour}{yellow!50!red}
\colorlet{commandcolour}{blue!60!black}
\colorlet{numpycolour}{blue!60!green}
\colorlet{literatecolour}{magenta!90!black}
\colorlet{promptcolour}{green!50!black}
\colorlet{specmethodcolour}{violet}

\usepackage{textcomp} 
\usepackage{listings}
\makeatletter
\lstdefinestyle{mylststyle}{
  basicstyle=%
    \ttfamily
    \lst@ifdisplaystyle\footnotesize\fi
}
\makeatother
\usepackage[%
  colorlinks  = true,%
  allcolors   = black,%
  urlcolor    = black,%
  pdfencoding = auto,%
  hyperindex,%
  pageanchor,%
  pdfdisplaydoctitle,%
  psdextra,%
  frenchlinks=true,
  final%
]{hyperref}
\usepackage{doi}

\usepackage[headsepline]{scrlayer-scrpage}
\clearpairofpagestyles
\setkomafont{pageheadfoot}{\footnotesize \rmfamily}
\setkomafont{pagination}{\footnotesize \rmfamily}
\setkomafont{title}{\linespread{1} \huge \rmfamily \textbf}
\setkomafont{subtitle}{\linespread{1} \large \rmfamily \textbf}
\setkomafont{author}{{\linespread{1} \rmfamily}}
\setkomafont{date}{\rmfamily}
\setkomafont{dedication}{{\linespread{1} \rmfamily}}
\setkomafont{publishers}{\rmfamily}
\setkomafont{subject}{\rmfamily}
\setkomafont{titlehead}{\rmfamily}
\addtokomafont{disposition}{\linespread{1} \rmfamily}

\DeclareNewLayer[%
  headsep,
  foreground,
  contents={%
    \begin{minipage}[t]{\textwidth}%
      \usekomafont{titlehead}{\csname @titlehead\endcsname\par}%
    \end{minipage}\par}
]{titlehead}
\DeclarePageStyleByLayers{titlepage}{titlehead}
\AddToHook{cmd/maketitle/after}{\thispagestyle{titlepage}}

\makeatletter
\def\@listI{\leftmargin\leftmargin
            \parsep 1\p@ \@plus2\p@ \@minus\p@
            \topsep 8\p@ \@plus2\p@ \@minus4\p@
            \itemsep1\p@ \@plus2\p@ \@minus\p@}
\makeatother

\newcommand*{\mainauthor}[1]{\def\MainAuthor{#1}}

\newcommand*{\headtitle}[1]{\def\HeadTitle{#1}}
\newcommand*{\maintitle}[1]{\def\MainTitle{#1}}
\newcommand*{\subt}[1]{\def\SubTitle{#1}}
\newcommand*{\fulltitle}[1]{\def\FullTitle{#1}}

\newcommand*{\orcid}[1]{\href{#1}{\includegraphics[height=10pt]{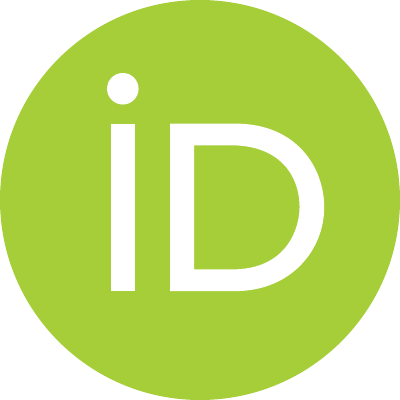}\,\nolinkurl{#1}}}
\newcommand*{\orcidlogo}[1]{\href{#1}{\includegraphics[height=10pt]{orcidlogo.pdf}}}

\newcommand*{\email}[1]{\href{mailto:#1}{\nolinkurl{#1}}}

\newcommand*{\pagestart}[1]{\def\Page{#1}}
\newcommand*{\keywords}[1]{\def\Keywords{#1}}

\ofoot[]{\pagemark}

\date{}

\hypersetup{
  pdfsubject  = {Proceedings of the 10th bwHPC Symposium, Freiburg September 2024},
  pdfkeywords = {bwHPC, HPC, 10th bwHPC Symposium},
}

\mainauthor{Daniel Jaschke}
\maintitle{Benchmarking Quantum Red TEA on CPUs, GPUs, and TPUs}
\subt{Submission to the 10$^{\mathrm{th}}$ bwHPC Symposium}
\fulltitle{\MainTitle{ }--{ }\SubTitle}
\headtitle{\MainTitle}
\pagestart{1}

\newcommand{\edigit}[2]{\textbf{#1}\textcolor{red}{#2}}

\newcommand{\plotlabel}[1]{\guillemotleft{}#1\guillemotright{}}
\newcommand{\numthreads}{n_{\mathrm{\scriptscriptstyle T}}}
\newcommand{\speedup}{S}

\newcommand{\applib}{\emph{\ref{app:i}}}
\newcommand{\apprgt}{\emph{\ref{app:ii}}}
\newcommand{\appmix}{\emph{\ref{app:iii}}}
\newcommand{\apppar}{\emph{\ref{app:iv}}}
\newcommand{\appgpu}{\emph{\ref{app:v}}}
\newcommand{\appchi}{\emph{\ref{app:vi}}}

\newcommand{\subfiga}{\includegraphics[height=0.2cm]{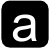}}
\newcommand{\subfigb}{\includegraphics[height=0.2cm]{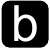}}

\newcommand{\orciddaniel}{https://orcid.org/0000-0001-7658-3546}
\newcommand{\orcidsimone}{https://orcid.org/0000-0002-8882-2169}
\newcommand{\orcidmarco}{https://orcid.org/0000-0002-3215-3453}
\newcommand{\orcidnora}{https://orcid.org/0000-0002-9490-2536}
\newcommand{\orcidluka}{https://orcid.org/0000-0002-9800-9200}

\newcommand{\numpy}{NumPy}
\newcommand{\numpycupy}{NumPy/CuPy}
\newcommand{\pytorch}{PyTorch}
\newcommand{\tensorflow}{TensorFlow}
\newcommand{\jax}{JAX}
\newcommand{\quantumtea}{Quantum TEA}
\newcommand{\quantumtealeaves}{Quantum TEA Leaves}
\newcommand{\quantumredtea}{Quantum Red TEA}
\newcommand{\qredtea}{Quantum Red TEA} 
\newcommand{\qtealeaves}{Quantum TEA Leaves} 

\newcommand{\hpccluster}{\emph{leonardo} (CINECA)}
\newcommand{\cpunode}{\emph{Intel Xeon Platinum 8480+}}

\newcommand{\gpunode}{\emph{NVIDIA A100-SXM-64GB} mounted on an \emph{Intel Xeon Platinum 8358}}
\newcommand{\gpumem}{64GB}

\newcommand{\cpuspeedup}{34}
\newcommand{\cpubestlib}{\pytorch{}}
\newcommand{\cpuchoices}{4 threads, SSSSDD sweep pattern, skip ERGT}
\newcommand{\gpuspeeduptotal}{94}
\newcommand{\gpuspeedupbestcpu}{2.76}

\newcommand{\tpunode}{TPU v4-8}
\newcommand{\tpuyear}{2021}

\newcommand{\uulm}{Institute for Complex Quantum Systems, 
  Ulm University,
  Albert-Einstein-Allee 11, 89069 Ulm, Germany}
\newcommand{\unipd}{Dipartimento di Fisica e Astronomia "G. Galilei" \& Padua Quantum Technologies Research Center,
 Universit{\`a} degli Studi di Padova, Italy I-35131, Padova, Italy}

\newcommand{\pdinfn}{INFN, Sezione di Padova, via Marzolo 8, I-35131,
  Padova, Italy}

\begin{document}

\begin{titlepage}
\thispagestyle{myheadings}
\raggedright
\vspace*{.05\textheight}

\begin{center}
{\usekomafont{title}
\MainTitle 
\par}
\end{center}

\vspace{\baselineskip}
{\usekomafont{subtitle}
\SubTitle{} 
\par}

\vspace{\baselineskip}

{\usekomafont{author}
\hyperref[sec:cauthor]{\MainAuthor{}$^{\mathrm{a,b,c}}$}
    \,\orcidlogo{\orciddaniel},
Marco Ballarin$^{\mathrm{b,c}}$%
    \,\orcidlogo{\orcidmarco},
Nora Reini{\'c}$^{\mathrm{b,c}}$%
    \,\orcidlogo{\orcidnora},
Luka Pavešić$^{\mathrm{b,c}}$%
    \,\orcidlogo{\orcidluka},
Simone Montangero$^{\mathrm{a,b,c}}$%
    \,\orcidlogo{\orcidsimone}
\par}

\vspace{.5\baselineskip}
{\usekomafont{author}
\small
$^{\mathrm{a}}$ \uulm{} \\
$^{\mathrm{b}}$ \unipd{} \\
$^{\mathrm{c}}$ \pdinfn{}
\par}

\vspace{\baselineskip}
\let\endtitlepage\relax
\end{titlepage}



\rohead{\HeadTitle} 
\lehead{\MainAuthor~et~al.} 

\hypersetup{
  pdftitle={\FullTitle},
  pdfauthor={\MainAuthor~et~al.},
  pdfkeywords={\Keywords},
}

\setcounter{page}{\Page}



\section*{Abstract}

We benchmark simulations of many-body quantum systems on heterogeneous hardware platforms
using CPUs, GPUs, and TPUs. 
We compare different linear algebra backends, e.g., 
\numpy{} versus the \pytorch{}, \jax{}, or \tensorflow{} libraries, as well as a mixed-precision-inspired approach
and optimizations for the target hardware.
\emph{\quantumredtea{}} out of the \emph{\quantumtea{}} library specifically addresses handling tensors with different libraries or hardware, where the tensors are the building blocks of tensor network algorithms.
The benchmark problem is a variational search of a ground state in an interacting model. This is a ubiquitous problem in quantum many-body physics, which we solve using tensor network methods. 
This approximate state-of-the-art method compresses quantum
correlations which is key to overcoming the exponential growth of the Hilbert space as a function of
the number of particles.  We present a way to obtain speedups of a factor of \cpuspeedup{}
when tuning parameters on the CPU, and an additional factor of \gpuspeedupbestcpu{} on top of
the best CPU setup when migrating to GPUs.

\pagebreak

\section{Introduction}                                                          \label{sec:intro}

Tensor network (TN) methods used for the simulation of quantum many-body systems
rely heavily on HPC resources. Carefully optimizing and benchmarking these algorithms
can result in significant savings of resources. This is essential
to emulate bigger systems, include more details in the description of the system,
or increase precision. Today's heterogeneous computing platforms open even 
more possibilities for optimization and a combination of various approaches.
We demonstrate where
computational gains are possible.
As using various hardware requires flexibility, we implement a software
design with interchangeable numerical libraries.

Tensor network methods have long been established as a method for the simulation 
of many-body quantum systems, see Refs.~\parencite{Schollwoeck2011,Orus2014,Banuls2022}. 
Today, they are used to numerically solve problems in fields ranging from condensed matter physics and 
lattice gauge theories and quantum simulation, to quantum circuit design~\parencite{montangero2018}. 
Quantum-inspired methods employ TNs to solve optimization problems,
partial differential equations, or
machine learning tasks~\parencite{Lucas2014,Gourianov2022,Stoudenmire2018}. The variety of applications
and their computational cost make TN methods an excellent target for exploring optimizations and
benchmarking.

We benchmark the \emph{\quantumtea{}} (Quantum Tensor network Emulator Applications)
library~\parencite{Silvi2017,qtealeaves_1_2_30,qredtea_0_0_15} with a ground state search for
a two-dimensional quantum Ising model~\parencite{SachdevQPT}.
\emph{\quantumtealeaves{}} implements various TN algorithms as a Python package.
\emph{\quantumredtea{}} enables additional tensor classes.
We choose the groundstate search of the two-dimensional quantum Ising model on a $16 \times 16$ lattice as
the benchmark problem, because of its relevance for the current research.
The tensor contractions and linear
algebra decompositions are either solved with \numpycupy{}, \pytorch{}, \jax{}, or
\tensorflow{}~\parencite{Numpy2020,Cupy2017,PyTorch2024,Jax2018,Tensorflow2015}.
In total, we outline seven approaches where we expect a potential speedup.
We successfully demonstrate how to leverage the different libraries, e.g.,
we gain a speedup factor of \cpuspeedup{} in the CPU benchmark.

The manuscript is organized as follows. In Sec.~\ref{sec:methods}, we introduce
TN methods to the extent necessary to follow the benchmarks
and describe the setup.
In Sec.~\ref{sec:benchmark}, we evaluate the
benchmark results for CPUs, GPUs, and TPUs.
We conclude in Sec.~\ref{sec:concl}.

\section{Methods and benchmark setup}                                           \label{sec:methods}

\begin{figure}[t]
  \begin{center}
    \begin{minipage}{0.49\linewidth}
      \begin{overpic}[width=0.95 \columnwidth,unit=1mm]{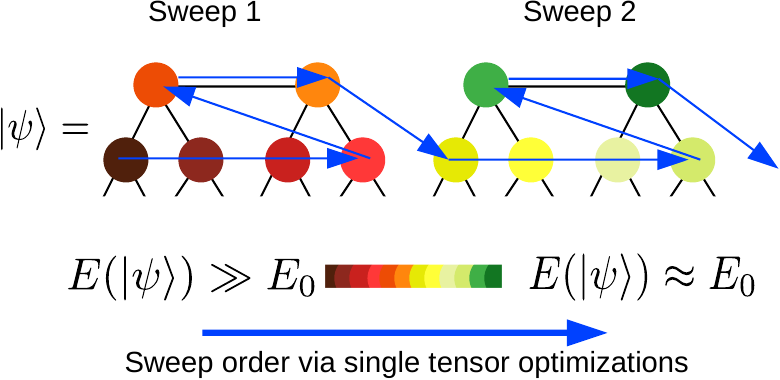}
        \put( 0,53){\subfiga}
      \end{overpic}
    \end{minipage}\hfill
    \begin{minipage}{0.49\linewidth}
      \begin{overpic}[width=0.95 \columnwidth,unit=1mm]{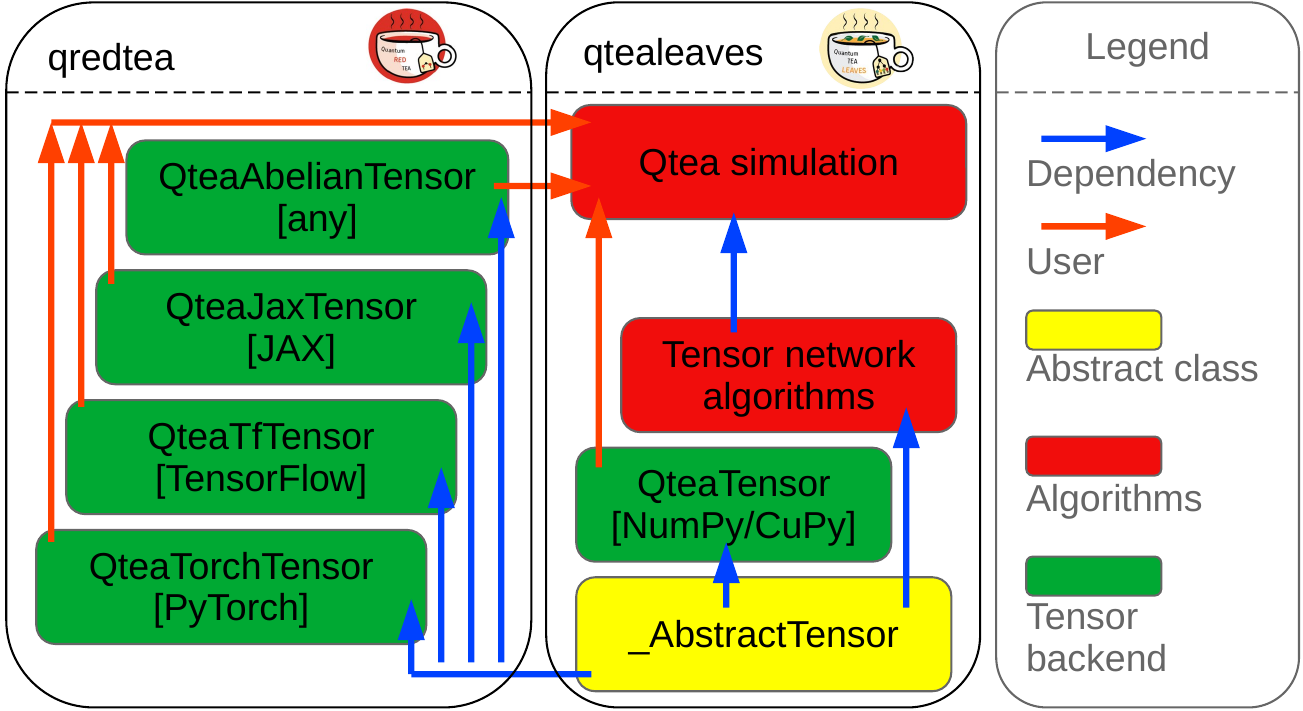}
        \put( 0,56){\subfigb}
      \end{overpic}
    \end{minipage}
    \caption{\emph{Algorithm and simplified library setup.}
      \subfiga{}~A binary tree tensor network represents the wavefunction $\ket{\psi}$ and the
      energy $E(\ket{\psi})$ is minimized via a series of sweeps, where each sweep consists
      of local tensor optimizations. With each optimization, the wavefunction
      converges towards the ground state with its ground state energy $E_{0}$.
      \subfigb{}~We consider two python packages from \emph{\quantumtea{}} for the ground state search.
      The \qtealeaves{} library comes with the tensor network algorithms
      as well as with an abstract class for tensors; moreover, the
      default tensors are implemented with \numpycupy{}. Any
      implementation of the abstract tensor class is suitable for the algorithms.
      The \qredtea{} library leverages the abstract class to implement optional
      tensors using \pytorch{}, \jax{}, and \tensorflow{}. Moreover, \qredtea{}
      encodes block-sparse tensors with an Abelian symmetry.
                                                                    \label{fig:qtea}}
  \end{center}
\end{figure}

The benchmark is set up around three main choices: (a)~we choose a binary tree tensor network (TTN) representation
for the wavefunction $\ket{\psi}$; (b)~we consider a ground state search, i.e., minimizing the energy $E$ of the wavefunction $\ket{\psi}$;
(c)~the underlying model is a two-dimensional quantum Ising model with
nearest-neighbor interactions. 
We discuss the rationale behind and consequences of the choices below.

In short, TN methods are based on efficiently, but approximately, representing
states of quantum systems, i.e., vectors living in a given Hilbert space, as products
of a set of tensors. Didactically, the approach separates a large vector into a set of
tensors with singular value decomposition, where only a set of states
associated with the largest singular values is kept. Due to an intrinsic
relation between entanglement, i.e., quantum correlation, and the distribution
of the singular values,
this is an optimal way to approximate quantum states.  
The number of kept singular values, i.e., the maximal size of a tensor's link,
uniquely determines the accuracy of the approximation. The \emph{maximal
bond dimension} $\chi$ is thus a central parameter of TN algorithms. 

We choose the TTN as it is particularly apt for representing quantum systems
with long-range interactions. These long-range interactions arise in the 2D quantum Ising model when
it is mapped into a 1D chain. The physical sites of the lattice are represented
as the leaves of the tree, while the rest of the network is there to transfer
entanglement between the distant sites. For a more detailed discussion of TN
methods, we direct the reader to \parencite{Schollwoeck2011}, while specifics
of TTNs are reviewed in \parencite{Silvi2017}.

To find the ground state of a given model, we iterate, i.e., we \emph{sweep}, through
the network and locally optimize each tensor in a way that minimizes the
energy. This procedure involves contracting the tensor with its environment, locally
solving an eigenproblem to find the smallest eigenvalue, and performing
decompositions to update the neighboring tensors afterwards.  
The ground state is reached by repeating this process multiple times, until
convergence is reached. See Fig. \ref{fig:qtea}a) for a sketch of an 8-qubit
TTN and its first two sweeps. The computational workload thus dominantly
comes from a large number of linear algebra tasks and manipulations of the
tensors: tensor contractions, sparse eigenproblems, as well as SVD, QR, and
eigen-decompositions.  Because the TTN uses rank-three tensors of dimensions
up to $\chi \times \chi \times \chi$, the complexity of these operations
scales as $\mathcal{O}(\chi^4)$. The simplified sketch of the library is shown
in Fig.~\ref{fig:qtea}b) with the two python modules, one abstract tensor class,
five different tensor backends, i.e., implementations of the abstract tensor class,
as well as the algorithms and simulations using the tensor classes.

We choose the ground state search as it is a ubiquitous problem in quantum physics,
but also because it leads to an easily accessible interpretation of the benchmark. 
Because the ground state energy $E_0$ is a global minimum of energy, 
lower energies represent a better outcome of the algorithm.
Comparison of computational time versus obtained energy is our main figure of merit.

The Hamiltonian $H$ for the ground state search of the quantum Ising model
in 2d is
\begin{align}                                                                 \label{eq:ham}
    H &= -J  \sum_{\langle i, j \rangle} \sigma_{i}^{x} \sigma_{j}^{x} -g \sum_{i} \sigma_{i}^{z} \, ,
\end{align}
where sites are labeled with $i = (i_{x}, i_{y})$ and $j$ referring to positions
in a 2d lattice and $\langle i, j \rangle$ contains all pairs of nearest-neighbors,
The Pauli matrices $\sigma^{x}$ and $\sigma^{z}$ define
the interactions with coupling $J=1$ and the local field with coupling $g$.
The linear system size is denoted by $N$, so that the total number of qubits is $N^{2}$,
In the following, we set $N=16$.
The ratio $g/J$ is set so that the system is close to the quantum critical point.
In this point, the ground state is strongly correlated, 
which imposes sufficient computational challenges.

We anticipate the potential for optimization in the following choices:

\begin{enumerate}[label=\roman*)]
    \item\label{app:i} \emph{Picking linear algebra libraries:}
      \numpy{} is the defacto-standard for numerical computation in python
      and is therefore the standard backend in \emph{\qtealeaves{}}.
      However, the alternatives to \numpycupy{} provide all necessary interfaces
      in terms of tensor contractions and linear algebra decompositions
      to run the ground state search. Our implementation approach
      relies on an abstract tensor class, which allows us to easily 
      run the same 
      TN algorithm with \numpycupy{}, \pytorch{}, \jax{},
      and \tensorflow{}. We provide an initial comparison between
      these libraries for \emph{\quantumtea{}} and its TTN ground state search.
      Note that the different libraries introduce different dependencies
      for low-level linear algebra, i.e., BLAS/LAPACK versus Eigen or
      Arpack versus a custom Lanzos solver.
    \item\label{app:ii} \emph{Exact renormalization group tensors (ERGT):}
      Each tensor in the lowest layer of the TTN contains two physical
      sites, which contain $d$ degrees of freedom (in our case $d=2$).
      These tensors thus have dimensions at most $d \times d \times d^2$,
      and can be represented exactly. 
      Written as a matrix $(d \times d) \times (d^2)$,
      it is an exact unitary transformation with a complete
      set of orthonormal vectors. Depending on the maximum bond
      dimension, more ERGTs exist in lower layers.
      Any optimization on the ERGTs can be accounted for
      in their parent tensors as no relevant degree of freedom
      has been truncated yet.
      We explore if skipping the ERGTs leads to a speedup.
    \item\label{app:iii} \emph{Mixed-precision tailored to TN methods:}
      the algorithm starts in a random guess state, which is far from
      optimum. Multiple iterations optimize the TN until
      it approximates the target state as good as possible with the given parameters of the algorithm.
      As arithmetic for single precision is faster than double precision, we exploit this
      by executing the first iterations in single-precision before
      moving the last sweeps to double-precision numbers.
    \item\label{app:iv} \emph{Parallelization on CPUs:}
      We investigate the parallel speedup of one simulation on CPUs for
      the different backend libraries. Parallelization occurs
      at the level of the BLAS/LAPACK or Eigen library.
    \item\label{app:v} \emph{GPUs-support:}
      The benefit of GPUs is well established and all
      four linear algebra libraries support GPUs. We benchmark simulations
      on CPU versus GPU.
    \item\label{app:vi} \emph{Optimization of tensor dimensions:}
      GPUs reach peak performance at matrix dimensions which respect
      the underlying hardware, e.g., rows of a matrix being a multiple
      of 128 bytes. We study whether enforcing
      such a dimension benefits the execution time of the complete
      algorithm as it does for matrix-matrix multiplications~\parencite{nvidiaGEMM}.
    \item\label{app:vii} \emph{TPUs support for simulations:}
      TPUs are another class of accelerators with
      features such as just-in-time compilation of functions for the best performance.
      The first step towards profiting from the platform is the migration
      towards the hardware, which we achieve by leveraging our different
      backend libraries.
\end{enumerate}

Together, the approaches encode the heterogeneous computing platforms
from various aspects including hardware and precision. The
benchmark executes on an \cpunode{} node or with a
\gpunode{} on \hpccluster{}. The TPU benchmark runs on Google's \tpunode{}.

\section{\quantumredtea{} benchmark}                                           \label{sec:benchmark}

The first part of the benchmark in Sec.~\ref{sec:cpu} uses CPUs where we analyze
higher-level optimization, like the mixed-precision approach tailored to the
ground state search. Based on the results, we look into the speedup that
can be gained for GPUs in Sec.~\ref{sec:gpu}. The number of simulation
parameters is too large to consider all combinations, so we in general focus
on one aspect of the simulation and pick one data point to compare
with the next optimization.

\subsection{CPU benchmark}                                                      \label{sec:cpu}

In the following section, we establish the baseline for further
optimization. We also use the opportunity to show the scaling of the simulation
time with the bond dimension $\chi$. The bond dimension $\chi$ directly influences
the underlying matrix dimension for the linear algebra operations.
The Hamiltonian of a quantum system can be complex in general, although many
models, including the one in Eq.~\eqref{eq:ham} can be represented with real-valued
Hamiltonians. Typically, simulations run with double-complex by default to capture
complex Hamiltonians; the mixed-precision implementation requires flexibility in the
data types, i.e., single-real, double-real, single-complex,
and double-complex are available, and we also have the opportunity to compare real
versus complex arithmetic.

\begin{figure}[t]
  \begin{center}
    \begin{minipage}{0.49\linewidth}
      \begin{overpic}[width=0.95 \columnwidth,unit=1mm]{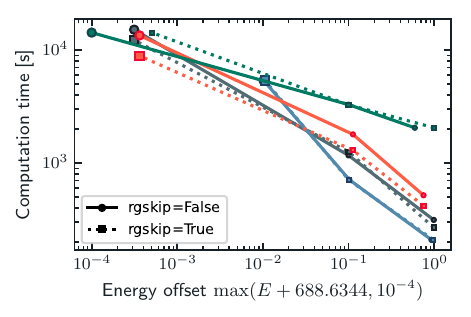}
        \put( 0,65){\subfiga{}}
        \put( 11,64){\includegraphics[width=0.8 \columnwidth]{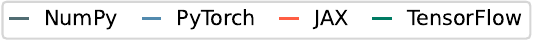}}
        \put(108,65){\subfigb{}}
      \end{overpic}
    \end{minipage}\hfill
    \begin{minipage}{0.49\linewidth}
    \tiny
    \begin{tabular}{@{} ccccc @{}}
      \toprule
      Sweeps & \numpy{} & \pytorch{} & \jax{} & \tensorflow{} \\
      \cmidrule[\heavyrulewidth](r){1-1} \cmidrule[\heavyrulewidth](rl){2-2} \cmidrule[\heavyrulewidth](rl){3-3} \cmidrule[\heavyrulewidth](rl){4-4} \cmidrule[\heavyrulewidth](l){5-5}
      SSSSSS&t =  977s & 627s &1365s &1914s \\
      &E = \edigit{-688.6}{2346} &\edigit{-688.6}{2502} &\edigit{-688.6}{0703} &\edigit{-688.}{59615} \\
      \cmidrule(r){1-1} \cmidrule(rl){2-2} \cmidrule(rl){3-3} \cmidrule(rl){4-4} \cmidrule(l){5-5}
      SSSSDD&t = 2249s & 976s &2198s &2721s \\
      &E = \edigit{-688.6}{2380} &\edigit{-688.63446}{} &\edigit{-688.634}{20} &\edigit{-688.6}{2307} \\
      \cmidrule(r){1-1} \cmidrule(rl){2-2} \cmidrule(rl){3-3} \cmidrule(rl){4-4} \cmidrule(l){5-5}
      SSSDDD&t = 3249s &1515s &2352s &3101s \\
      &E = \edigit{-688.6}{2382} &\edigit{-688.6}{2908} &\edigit{-688.634}{21} &\edigit{-688.63}{315} \\
      \cmidrule(r){1-1} \cmidrule(rl){2-2} \cmidrule(rl){3-3} \cmidrule(rl){4-4} \cmidrule(l){5-5}
      DDDDDD&t = 5046s &2031s &3394s &4050s \\
      &E = \edigit{-688.63}{382} &\edigit{-688.63}{394} &\edigit{-688.6}{2370} &\edigit{-688.63}{370} \\
      \cmidrule(r){1-1} \cmidrule(rl){2-2} \cmidrule(rl){3-3} \cmidrule(rl){4-4} \cmidrule(l){5-5}
      ZZZZZZ&t = 12207s &5379s &8881s &14233s \\
      &E = \edigit{-688.634}{13} &\edigit{-688.6}{2395} &\edigit{-688.634}{08} &\edigit{-688.63}{394} \\
      \bottomrule
    \end{tabular}
    \end{minipage}
    \caption{\emph{CPU benchmark: baseline and mixed-precision tailored to tensor networks.}
      \subfiga{}~The baseline for the ground state search benchmark with no
      optimizations is given by the data with \plotlabel{rgskip=False}. To allow for log-error plots,
      we plot $\max\left( E - \min_{\chi, \mathrm{lib}, \mathrm{rgskip}} (E(\cdot)), \epsilon=10^{-4}\right)$.
      The largest markers refer to the largest bond dimension out of $\chi \in \{16, 32, 64 \}$
      for each backend library, where the different backend libraries perform differently.
      Then, we tune our setting and skip small exact tensors in the optimization, see \plotlabel{rgskip=True},
      leading only to a speedup for \jax{}.
      \subfigb{}~Mixed-precision allows to increase precision as we converge to the ground state in multiple sweeps.
      We perform six sweeps with single-real (S), double-real (D), and double-complex (Z) data types
      and show selected entries. Mixed-precision always leads to a speedup;
      note that precision decreases for \numpy{} and \tensorflow{} with more single-precision sweeps,
      while the precision fluctuates for \pytorch{} and \jax{}.
                                                                    \label{fig:benchmark_cpu}}
  \end{center}
\end{figure}

In Fig.~\ref{fig:benchmark_cpu}a), we address the effects of points \applib{} and
\apprgt{} for different bond dimensions; here, we start with a complex-valued ansatz.
We compare the obtained energy with computation time.
To enable log-log plots, we plot the energy relative to the lowest energy obtained across all simulations $E_\mathrm{min}$ and manually set the best simulation's error to the next smaller power of ten, here $10^{-4}$.
We expect simulations with higher bond dimensions to achieve better
precision at the cost of a longer computation time;
the diagonal trend in Fig.~\ref{fig:benchmark_cpu}a) reflects the improvement with bond dimension.
We use $\chi \in \{16, 32, 64 \}$, with the size of the marker corresponding to the bond dimension used.
Surprisingly, we find a difference in runtime of a factor of two between libraries, e.g., \pytorch{} versus \numpy{} at $\chi = 64$.
We find that the order from the fastest to slowest at $\chi = 64$ is: \jax{}, \tensorflow{}, \pytorch{}, and \numpy{}. 
Our results demonstrate the advantage of choosing the most suitable backend-library.

The speedup gained by skipping exactly represented tensors in optimization sweeps, see point \apprgt{}, is also 
shown in Fig.~\ref{fig:benchmark_cpu}a): the data set with squared markers and dashed lines skips small tensors. 
The impact depends on the choice of library and $\chi$, from no clear impact
for \numpy{}, \pytorch{}, and \tensorflow{}, to a slight improvement in runtime and precision for \jax{}. 

In Fig.~\ref{fig:benchmark_cpu}b), we investigate the effect of mixed-precision 
by performing six sweeps while changing the precision.
The label S denotes a sweep with single-real precision, and D with double-real precision.
The final obtained energy total wall-time ought to be compared to six sweeps in double-complex, label \plotlabel{ZZZZZZ}.
The simulation time decreases for more single-precision sweeps;
the speedup comes with losing the precision of the ground state
energy for \numpy{} and \tensorflow{}. In contrast, the precision fluctuates for \pytorch{}
and \jax{} and no clear dependency with the number of single-precision sweeps is visible.
For example, we find that the \numpy{} backend loses one digit of precision for an SSSSDD pattern.
These results show that it is possible to gain a speedup at a reasonable precision
by choosing a suitable strategy for approach \appmix{}.

\begin{figure}[t]
  \begin{center}
    \begin{minipage}{0.49\linewidth}
    \tiny
    \begin{tabular}{@{} ccccc @{}}
      \toprule
      Cores & \numpy{} & \pytorch{} & \jax{} & \tensorflow{} \\
      \cmidrule[\heavyrulewidth](r){1-1} \cmidrule[\heavyrulewidth](rl){2-2} \cmidrule[\heavyrulewidth](rl){3-3} \cmidrule[\heavyrulewidth](rl){4-4} \cmidrule[\heavyrulewidth](l){5-5}
      1 & t = 2249s &  976s & 2198s & 2721s \\
      &E = \edigit{-688.6}{2380} & \edigit{-688.63446}{} & \edigit{-688.634}{20} & \edigit{-688.6}{2307} \\
      \cmidrule(r){1-1} \cmidrule(rl){2-2} \cmidrule(rl){3-3} \cmidrule(rl){4-4} \cmidrule(l){5-5}
      4 & t = 5424s &  443s & 1727s & 2619s \\
      &E = \edigit{-688.6}{2424} & \edigit{-688.6}{2393} & \edigit{-688.63}{393} & \edigit{-688.634}{21} \\
      \cmidrule(r){1-1} \cmidrule(rl){2-2} \cmidrule(rl){3-3} \cmidrule(rl){4-4} \cmidrule(l){5-5}
      14 & t = 5486s & 1680s & 1426s & 2624s \\
      &E = \edigit{-688.6}{2397} & \edigit{-688.6}{2391} & \edigit{-688.63}{393} & \edigit{-688.6}{2385} \\
      \cmidrule(r){1-1} \cmidrule(rl){2-2} \cmidrule(rl){3-3} \cmidrule(rl){4-4} \cmidrule(l){5-5}
      16 & t = 6099s & 1523s & 1372s & 2281s \\
      &E = \edigit{-688.6}{2315} & \edigit{-688.63}{022} & \edigit{-688.63}{393} & \edigit{-688.6}{2376} \\
      \cmidrule(r){1-1} \cmidrule(rl){2-2} \cmidrule(rl){3-3} \cmidrule(rl){4-4} \cmidrule(l){5-5}
      112 & t = 4475s & 2756s & 1081s & 2130s \\
      &E = \edigit{-688.6}{2380} & \edigit{-688.6}{2917} & \edigit{-688.63}{393} & \edigit{-688.6}{2378} \\
      \bottomrule
    \end{tabular}
    \end{minipage}\hfill
    \begin{minipage}{0.49\linewidth}
      \begin{overpic}[width=0.95 \columnwidth,unit=1mm]{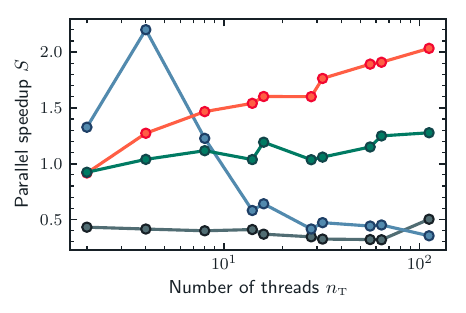}
        \put( 0,65){\subfigb{}}
        \put( 12,64){\includegraphics[width=0.8 \columnwidth]{legend_a.pdf}}
        \put(-110,65){\subfiga{}}
      \end{overpic}
    \end{minipage}
    \caption{\emph{Built-in CPU parallelization of tensor operations via the linear algebra libraries.}
      \subfiga{}~The energies show no significant trend when changing the number
      of threads. The wall times can decrease moderately for \pytorch{}, \jax{}, and \tensorflow{}.
      In contrast, parallelization even slows down simulations with the \numpy{}-backend
      as well as for \pytorch{} with many cores.
      \subfigb{}~The parallel speedup $\speedup$ over the number of threads $\numthreads$
      reflects the data of the table in a) with parallel speedup except for \numpy{}.
      The use of multiple CPU is not justified by the observed speedup for any backend,
      except for \pytorch{} with two and four cores.
                                                                    \label{fig:benchmark_parallel}}
  \end{center}
\end{figure}

As the last step for the CPU benchmark, we consider parallelization on a single node.
Numerical libraries like openBLAS come with threading; therefore, we
expect to observe a parallel speedup when increasing the number of threads.
No additional parallelization in our TN algorithms, e.g., of sweeps, is required to profit from this parallelization.
Figures~\ref{fig:benchmark_parallel}a) and b) indicate a parallel speedup $S = T_{\mathrm \scriptscriptstyle parallel} / T_{\mathrm \scriptscriptstyle single}$ for
the \pytorch{}, \jax{} and \tensorflow{} backend. \pytorch{} peaks at a speedup of about $2$ for 4 cores;
\jax{} and \tensorflow{} reach their maximum speedup at $112$ cores. Potential
explanations for the features in the speedup are optimizations for powers of 2
for \pytorch{}, see each local maxima at 8, 16, 32, and 64 threads, or the NUMA node
size of 14 for \jax{}, see local maxima for 14 or 28 threads.
The configuration
of \numpy{} seems ill-configured for the profile on leonardo. Overall, the parallel speedup
stays behind the number of cores and saturates for the given problem sizes
below 3. Additional tuning is necessary to leverage the resources and identify
bottlenecks.

To conclude, we find that optimizing approaches \applib{}
through \apppar{} provides a potential speedup, but the improvement depends on
picking the right library-backend in the first place. This applies to the question of parallelization in particular.
The overall speedup we gain is significant, with a factor of \cpuspeedup{} with the
\cpubestlib{}-backend for the optimized setup (\cpuchoices{}) in comparison to the original double-complex simulation with the \numpy{}-backend.

\subsection{GPU benchmark}                                                      \label{sec:gpu}

In the following, we first address the migration to the GPU, i.e.,
approach \appgpu{}, and then look into fine-tuning the tensor dimensions
as proposed in approach \appchi{}. 
We compare a single CPU core to a single GPU. Executing the
simulation on the GPU comes with some additional constraints and options. The
main constraint of the GPU is the \gpumem{} of memory, where we are not
exceeding this \gpumem{} for 256 sites and $\chi = 64$.
Moreover, we tune the preference of the linear algebra
algorithms according to our preliminary studies, e.g., solving
SVDs via eigendecompositions.
The settings of the algorithm require SVDs and truncated bond dimensions
are possible.

\begin{figure}[t]
  \begin{center}
    \begin{minipage}{0.49\linewidth}
      \begin{overpic}[width=0.95 \columnwidth,unit=1mm]{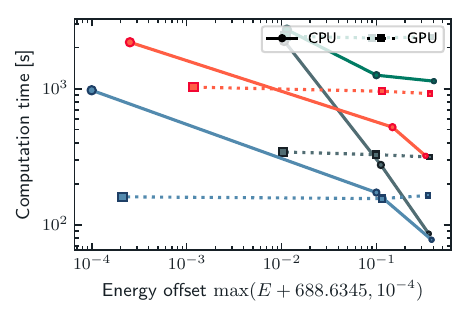}
        \put( 0,65){\subfiga{}}
        \put( 12,64){\includegraphics[width=0.8 \columnwidth]{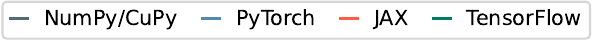}}
      \end{overpic}
    \end{minipage}\hfill
    \begin{minipage}{0.49\linewidth}
      \begin{overpic}[width=0.95 \columnwidth,unit=1mm]{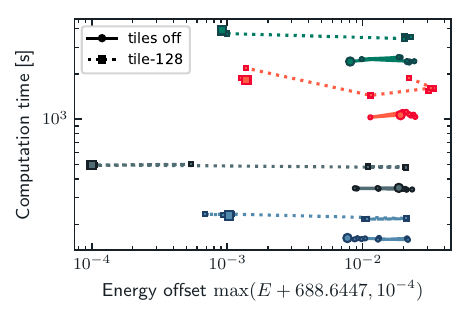}
        \put( 0,65){\subfigb{}}
        \put( 12,64){\includegraphics[width=0.8 \columnwidth]{legend_e.pdf}}
      \end{overpic}
    \end{minipage}
    \caption{\emph{GPU benchmark for mixed-precision and block-size optimization.}
      \subfiga{}~We port the mixed-precision simulations to the GPU and compare their speedup
      to their corresponding CPU version. We iterate over the bond dimensions $\chi \in \{ 16, 32, 64 \}$
      where the largest bond dimension in each curve is indicated with the bigger markers.
      \subfigb{}~Matrix-matrix multiplications are optimized in CUDA for the
      block-size of the memory, i.e., tiles are enforced internally. We can enforce
      our tensor shapes to be a multiple of this tile size and observe if there is
      additional speedup; but tiling leads to a slowdown for all backends.
      Each curve consists of bond dimension $\chi \in \{60, \ldots 68\}$ with
      the biggest bond dimension indicated by a bigger marker.
                                                                    \label{fig:benchmark_gpu}}
  \end{center}
\end{figure}

Figure~\ref{fig:benchmark_gpu}a) compares the energy obtained with different bond
dimensions $\chi$ and
different libraries on CPU and GPU. We use a sweep pattern of four single-precision
plus two double-precision sweeps with a real-valued ansatz. We see that for both
devices the \pytorch{}-backend is much faster while achieving equal or almost equal precision as other libraries. 
Note that the increase in computational time with increasing $\chi$ is much slower
for the GPU than for the CPU. 
We find that the GPUs are 
slower for very small $\chi$, but provide a sizeable speedup over CPU already
for $\chi=32$. As expected, approach \appgpu{} is successful and
leads to an additional speedup over the best CPU simulation of \gpuspeedupbestcpu{} times.

Now, we move to the tiling size, which potentially helps to improve performance
as demonstrated in Ref.~\parencite{nvidiaGEMM}. Figure~\ref{fig:benchmark_gpu}b)
shows two data sets for $\chi \in \{60, \ldots, 68 \}$ without and with enforcing
a tiling, solid versus dotted lines.
If activated, we enforce a tiling of 128 bytes; note that $\chi = 64$ enforces
only the maximal bond dimension, while tensors may have other bond dimensions due to
the truncation of singular values. We expect
a separation of the two data sets in case enforcing the tiling has an
effect, but the data does not support the claim. In contrast, the runtime
increases slightly when enforcing tiling on our side. 
Therefore, we conclude that \appchi{} provides no benefit.

In summary, the migration to the GPU as approach \appgpu{} is successful with
a total speedup of \gpuspeeduptotal{} over the double-complex \numpy{}-backend
baseline on CPUs. Further tuning as enforcing a tiling dimension does not
lead to an additional speedup.

\subsection{TPU benchmark}                                                      \label{sec:tpu}

We move to the TPU benchmark, which has as an accelerator similar challenges for
an implementation as GPUs, e.g., data transfer between host and device. \pytorch{},
\tensorflow{}, and \jax{} support TPUs. We focus on \jax{} after an initial assessment
and considering previous studies on linear algebra with
\jax{}~\parencite{Lewis2022}. \jax{} relies on just-in-time (jit) compilation of
functions that are cached for each data type, shape, etc. The jit compiler
opens up possibilities for optimization, e.g., avoiding unnecessary compilation.

\begin{figure}[t]
  \begin{center}
    \begin{minipage}{0.49\linewidth}
    \begin{overpic}[width=0.01 \columnwidth,unit=1mm]{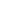}
        \put( 0,61){\subfiga{}}
      \end{overpic}
    \\
    \tiny
    \begin{tabular}{@{} cccc @{}}
      \toprule
      Bond dimension & CPU & XLA & XLA + tile=128 \\
      \cmidrule[\heavyrulewidth](r){1-1} \cmidrule[\heavyrulewidth](rl){2-2} \cmidrule[\heavyrulewidth](rl){3-3} \cmidrule[\heavyrulewidth](l){4-4} 
    $\chi = 32$ & t=1065s & 1131s & n.a.\\
    & E=\edigit{-688.}{51693} & \edigit{-68}{7.98895} & n.a.\\
      \cmidrule(r){1-1} \cmidrule(rl){2-2} \cmidrule(rl){3-3} \cmidrule(l){4-4}
    $\chi = 64$ & t=1823s & 1180s & 1625s\\
    & E=\edigit{-688.61092}{} & \edigit{-68}{7.06878} & \edigit{-68}{4.32439}\\
      \cmidrule(r){1-1} \cmidrule(rl){2-2} \cmidrule(rl){3-3} \cmidrule(l){4-4}
    $\chi = 128$ & t=4692s & 1244s & 1701s\\
    & E=\edigit{-688.}{57309} & \edigit{-68}{6.84771} & \edigit{-6}{68.38448}\\
      \bottomrule
    \end{tabular}
    \end{minipage}\hfill
    \begin{minipage}{0.49\linewidth}
      \begin{overpic}[width=0.01 \columnwidth,unit=1mm]{white_square.png}
        \put( 50,65){\subfigb{}}
      \end{overpic}
      \\
      \tiny
    \begin{tabular}{@{} cccc @{}}
      \toprule
      XLA compile log & CPU & XLA & XLA + tile=128 \\
      \cmidrule[\heavyrulewidth](r){1-1} \cmidrule[\heavyrulewidth](rl){2-2} \cmidrule[\heavyrulewidth](rl){3-3} \cmidrule[\heavyrulewidth](l){4-4} 
    Time $(\chi = 64$) & 41s & 32s & 48s\\
    Counter & 459 & 459 & 742\\
      \cmidrule(r){1-1} \cmidrule(rl){2-2} \cmidrule(rl){3-3} \cmidrule(l){4-4}
    Time $(\chi = 128$) & 42s & 56s & 140s\\
    Counter & 459 & 459 & 1207\\
      \bottomrule
    \end{tabular}

    \end{minipage}
    \caption{\emph{TPU benchmark for single-precision and block-size optimization with \jax{}.}
      \subfiga{}~We port the mixed-precision simulations to the TPU and compare
      different parameterizations. We observe approximately a factor $2$
      in difference in simulation time for doubling the bond dimension $\chi$. The two data points
      per line refers to four versus six single-precision steps with the bigger
      marker for six single-precision sweeps. Finally,
      we check if tiling bond dimensions to 128 bytes has any effect, e.g.,
      due to tiling or fewer jit-compiling efforts.
      \subfigb{}~We extract the number of jit-compile events and their cumulated
      time. A change in data types leads to about 20 more calls to the
      compiler. Higher bond dimensions lead to more compilation time.
      Tiling does not overcome this problem as truncations do not
      lead to different dimensions in our example.
                                                                    \label{fig:benchmark_tpu}}
  \end{center}
\end{figure}

We benchmark on Google's \tpunode{} for three bond dimensions $\chi \in \{32, 64, 128\}$, six
single-precision sweeps, and tiling on/off. The computation times of the \tpunode{} cannot compete
with GPUs and are closer to the ones of CPUs, see Fig.~\ref{fig:benchmark_tpu}a). At bond dimension $\chi = 64$, there is a speedup of the TPU over the CPU, albeit with a smaller precision.
The difference in wall time when doubling  the bond dimensions is less than
$1.1$ on the TPU, which is far below the expected scaling with $\mathcal{O}(\chi^4)$.
This means that other contributions still present a significant contribution in comparison to the
expensive linear algebra steps.

As each function has to be
compiled for each set of parameters, approach \appchi{} seems to fit into
the beneficial optimization for TPUs: limiting the number of possible bond dimensions
also limits the number of compile events. Surprisingly, it is not beneficial; computation
time increases, and the compile time of the jit-compiler also increases, because the number
of compile events increases. In figure~\ref{fig:benchmark_tpu}b), we extract
the actual number of jit-compiler calls and the compile time
to track the effects of bond dimension and tiling.  
The higher number
of jit-compilations for tiling can be explained by additional compilations
of functions only needed for tiling as a first hypothesis. An additional
look into the data shows further that simulations without tiling max out
their bond dimension and therefore encounter only a limited number of
combinations.

These results concerning the precision are not \textit{per se} a statement
against TPUs as the scaling of the computation with the bond dimension $\chi$ is promising.
Further optimization steps remain unexplored.
For example, removing trivial dimensions before jit-compiling
functions, enabling jited-functions in our library instead of completely relying
on the \jax{}-backend. The smallest size of the TPU used here also allows one to further scale it up as successfully shown for linear algebra in Ref.~\parencite{Lewis2022}. A comparison with newer hardware is another
future step; \tpunode{} has been released in \tpuyear{}.

\subsection{Block-sparse tensors for conserved symmetries}                      \label{sec:symm}

As the last part of the benchmark, we demonstrate that the abstract tensor
class extends into block-sparse tensors. These block-sparse tensors enable us
to handle quantum systems with conserved quantities. The Hamiltonian
in Eq.~\eqref{eq:ham} has a $\mathbb{Z}_{2}$ Abelian symmetry.
The block-sparse tensors rely on the abstract tensor in two 
ways. First, block-sparse tensors have to comply with the abstract tensor class so that
the TTN algorithms can employ them as tensors for a TN. Second, the block-sparse tensors have to contain a tensor for each block; these are dense tensors based on \numpycupy{}, \pytorch{}, \jax{}, or \tensorflow{}.

\begin{figure}[t]
  \begin{center}
    \begin{minipage}{0.49\linewidth}
      \begin{overpic}[width=0.95 \columnwidth,unit=1mm]{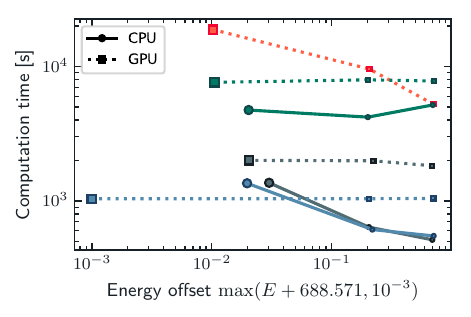}
        \put( 0,66){\subfiga{}}
        \put(8,64){\includegraphics[width=0.83 \columnwidth]{legend_e.pdf}}
      \end{overpic}
    \end{minipage}\hfill
    \begin{minipage}{0.49\linewidth}
      \begin{overpic}[width=0.95 \columnwidth,unit=1mm]{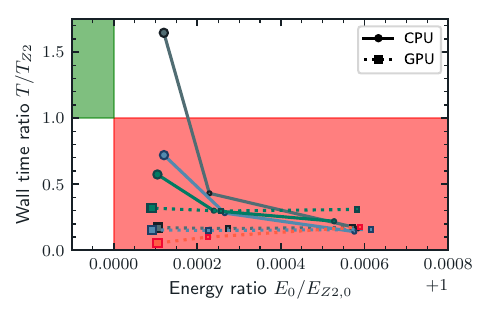}
        \put( 0,63){\subfigb{}}
        \put(5,61){\includegraphics[width=0.83 \columnwidth]{legend_e.pdf}}
      \end{overpic}
    \end{minipage}
    \caption{\emph{Block-sparse tensor benchmark for conserved symmetries.}
      \subfiga{}~We analyze the TTN with a conserved $\mathbb{Z}_{2}$ symmetry
         and its energies and runtimes for both CPU and GPU. As linear algebra
         operations are on smaller problems, GPUs do not yet show a speedup
         for $\chi \le 64$.
      \subfigb{}~We compare the data with and without symmetries via the
         ratios of the energy and wall time. Ratios in the red (green)
         rectangle are a decline (improvement) with respect to both
         energy and wall time. CPUs profit with respect to the wall
         time.
                                                                    \label{fig:benchmark_z2}}
  \end{center}
\end{figure}

In Fig.~\ref{fig:benchmark_z2}a), we show the computation time over the energy
of the ground state with symmetry
for different bond dimensions $\chi \in \{ 16, 32, 64 \}$.
In Fig.~\ref{fig:benchmark_z2}b) we compare to the same simulation 
without the $\mathbb{Z}_{2}$ symmetry. 
By taking the ratio of energies and wall times,
we see whether both aspects improved (see green rectangle) or whether both aspects
deteriorate (red rectangle). For $\chi = 64$, we find a 
speedup in computation time for CPUs with \numpy{} when using block-sparse tensors, although convergence is still slightly worse. 
The expected speedup of block-sparse tensors depends on the exploited symmetry.
Our Hamiltonian only splits into two blocks and
the upper bound for a speedup is 4. 
Because the matrix dimension for linear algebra operations is effectively smaller when using symmetric tensors, 
the GPU has less impact in comparison with the CPU for the bond dimensions
$\chi \le 64$ shown here.
To summarize, the concept of abstract classes extends to symmetric
tensors, however, parallel loops of decompositions are not exploited yet and have the potential for an additional speedup in the future.

\section{Conclusions}                                                           \label{sec:concl}

We have demonstrated a benchmark for tensor network algorithms used to simulate
many-body quantum systems. Due to the uprise of quantum-inspired methods, this
class of algorithms has raised interest beyond quantum systems, e.g., for optimization
or machine learning problems. We identify seven aspects that have the potential
to reduce computation time spanning different hardware, i.e., CPUs, GPUs, and TPUs, backend libraries, and precision of the underlying arithmetic.

When using CPUs, we find that the most important aspect is the choice of
the linear algebra library.
Additionally, using lower precision for initial iterations also provides a sizeable speedup. 
Combined with the other approaches like parallelization, this strategy leads to a
speedup factor of \cpuspeedup{} compared to the initial baseline.

For GPUs, we demonstrate an additional speedup of \gpuspeedupbestcpu{} for the \pytorch{}-backend
over the best CPU implementation (\cpuchoices{}). The overall speedup from the initial starting
point to the optimal GPU implementation is \gpuspeeduptotal{}.

We see further potential in matching the strengths of the linear algebra libraries 
with the type of hardware.  
An example is the TPU architecture, where we achieve a working example, but no significant speedup at sufficient precision yet. Extracting the best computational
performance requires in the future considering just-in-time compilation and rewriting functions with this aspect in mind. 
This approach then benefits both TPUs and GPUs.

\subsubsection*{Corresponding Author}                                           \label{sec:cauthor}

Daniel Jaschke:~\email{daniel-1.jaschke@uni-ulm.de}\\
Institute for Complex Quantum Systems, Ulm University, Germany

\subsubsection*{ORCID}

Daniel Jaschke\,\orcid{\orciddaniel}\\
Marco Ballarin\,\orcid{\orcidmarco}\\
Nora Reini{\'c}\,\orcid{\orcidnora}\\
Luka Pavešić,\orcid{\orcidluka}\\
Simone Montangero\,\orcid{\orcidsimone}\\

\subsubsection*{Data availability}

The scripts and source code are available via
Refs~\parencite{qtealeaves_1_2_30} and \parencite{qredtea_0_0_15}. We provide
the datasets and figures via Ref.~\parencite{qredtea_benchmark_data}.

\subsubsection*{Acknowledgements}

We thank Alessandro Lonardo, Francesco Simula, Ilaria Siloi, and Pietro Silvi for discussions and feedback.
We acknowledge financial support from the Italian Ministry of University and 
Research (MUR) via PRIN2022 project TANQU, and the Departments of Excellence
grant 2023-2027 Quantum Frontiers; from the European Union via H2020 projects
EuRyQa and textarossa, the QuantERA projects QuantHEP and T-NISQ, and the Quantum Flagship
project  Pasquans2, from the German Federal Ministry of Education and Research
(BMBF) via the funding program quantum technologies $-$ from basic research to
market $-$ project QRydDemo, and from the World Class Research Infrastructure
$-$ Quantum Computing and Simulation Center (QCSC) of Padova University.
N.~R. received support from the European Union via the UNIPhD programme
(Horizon 2020 under Marie Skłodowska-Curie grant agreement No. 101034319 and
NextGenerationEU).
We also acknowledge computation time supported support by the state of
Baden-Württemberg through bwHPC and the German Research Foundation (DFG)
through grant no INST 40/575-1 FUGG (JUSTUS 2 cluster),
on Cineca's \emph{leonardo} machine, the dibona cluster via the
project textarossa, and via Google's TPU Research Cloud program.

\printbibliography

@article{Banuls2022,
    author = {Ba\~{n}uls, Mari Carmen},
    title = "Tensor Network Algorithms: A Route Map",
    journal = "Annual Review of Condensed Matter Physics",
    volume = "14",
    number = "1",
    pages = "173-191",
    year = "2023",
    doi = "10.1146/annurev-conmatphys-040721-022705"
}

@article{Gourianov2022,
  author = {Gourianov, Nikita and Lubasch, Michael and Dolgov, Sergey and van den Berg, Quincy Y. and Babaee, Hessam and Givi, Peyman and Kiffner, Martin and Jaksch, Dieter},
  title = {A quantum-inspired approach to exploit turbulence structures},
  journal = {Nature Computational Science},
  year = {2022},
  month = {Jan},
  day = {01},
  volume = {2},
  number = {1},
  pages = {30-37},
  issn = {2662-8457},
  doi = {10.1038/s43588-021-00181-1},
}

@article{Lucas2014,
  author = {Lucas, Andrew },
  title = {Ising formulations of many NP problems},
  journal = {Frontiers in Physics},
  volumne = {2},
  year = {2014},
  doi = {10.3389/fphy.2014.00005},
  issn = {2296-424X}
}

@book{montangero2018,
    author = "Montangero, S.",
    title = "Introduction to Tensor Network Methods",
    publisher = "Springer Nature Switzerland AG",
    year = "2018",
    address = "Cham, CH",
    doi = "10.1007/978-3-030-01409-4"
}

@article{Orus2014,
    author = "Or\'us, Rom\'an",
    title = "{A practical introduction to tensor networks: {M}atrix product states and projected entangled pair states}",
    journal = "Annals of Physics",
    volume = "349",
    number = "",
    pages = "117 - 158",
    year = "2014",
    note = "",
    issn = "0003-4916",
    doi = "10.1016/j.aop.2014.06.013",
}

@software{qtealeaves_1_2_30,
  author = {Bacilieri, Davide and Ballarin, Marco and Cataldi, Giovanni and Costantini, Aurora and Jaschke, Daniel and Magnifico, Giuseppe and Montangero, Simone and Notarnicola, Simone and Pagano, Alice and Pavešić, Luka and Rattacaso, Davide and Rigobello, Marco and Reinić, Nora and Scarlatella, Simone and Silvi, Pietro and Wanisch, Darvin},
  title  = {Quantum TEA: qtealeaves},
  month  = {Jan},
  year = {2024},
  publisher  = {Zenodo},
  version  = {1.2.30},
  doi  = {10.5281/zenodo.13383350}
}

@software{qredtea_0_0_15,
    author = {Ballarin, Marco and Jaschke, Daniel and Montangero, Simone and Pavešić, Luka and Reinić, Nora},
    title = {Quantum TEA: qredtea},
    month = {Aug},
    year = {2024},
    publisher = {Zenodo},
    version  = {0.0.15},
    doi = {10.5281/zenodo.13385250}
}

@software{qredtea_benchmark_data,
  author = {Jaschke, Daniel and Ballarin, Marco and Reinić, Nora and Pavešić, Luka and Montangero, Simone},
  title = {{Simulation data and figures for "Benchmarking Quantum Red TEA on CPUs, GPUs, and TPUs"}},
  month = {Aug},
  year = {2024},
  publisher = {Zenodo},
  version = {v0.0.15},
  doi = {10.5281/zenodo.13386420}
}

@book{SachdevQPT,
  author = {Sachdev, Subir},
  title = {{Quantum Phase Transitions}},
  publisher = {Cambridge University Press},
  year = {2011},
  edition = {2nd},
  address = {Cambridge, United Kingdom},
  doi = {10.1017/CBO9780511973765}
}

@article{Schollwoeck2011,
    author = {Schollw\"{o}ck, Ulrich},
    title = "The density-matrix renormalization group in the age of matrix product states",
    journal = "Annals of Physics",
    volume = "326",
    number = "1",
    pages = "96 - 192",
    year = "2011",
    note = "{January 2011 Special Issue}",
    issn = "0003-4916",
    doi = "10.1016/j.aop.2010.09.012"
}

@article{Silvi2017,
    author = "Silvi, Pietro and Tschirsich, Ferdinand and Gerster, Matthias and Jünemann, Johannes and Jaschke, Daniel and Rizzi, Matteo and Montangero, Simone",
    title = "{The Tensor Networks Anthology: Simulation techniques for many-body quantum lattice systems}",
    journal = "SciPost Phys. Lect. Notes",
    pages = "8",
    year = "2019",
    publisher = "SciPost",
    doi = "10.21468/SciPostPhysLectNotes.8"
}

@article{Stoudenmire2018,
  doi = {10.1088/2058-9565/aaba1a},
  year = {2018},
  month = {Apr},
  publisher = {IOP Publishing},
  volume = {3},
  number = {3},
  pages = {034003},
  author = {E. Miles Stoudenmire},
  title = {Learning relevant features of data with multi-scale tensor networks},
  journal = {Quantum Science and Technology}
}

@inproceedings{PyTorch2024,
author = {Ansel, Jason and Yang, Edward and He, Horace and Gimelshein, Natalia and Jain, Animesh and Voznesensky, Michael and Bao, Bin and Bell, Peter and Berard, David and Burovski, Evgeni and Chauhan, Geeta and Chourdia, Anjali and Constable, Will and Desmaison, Alban and DeVito, Zachary and Ellison, Elias and Feng, Will and Gong, Jiong and Gschwind, Michael and Hirsh, Brian and Huang, Sherlock and Kalambarkar, Kshiteej and Kirsch, Laurent and Lazos, Michael and Lezcano, Mario and Liang, Yanbo and Liang, Jason and Lu, Yinghai and Luk, CK and Maher, Bert and Pan, Yunjie and Puhrsch, Christian and Reso, Matthias and Saroufim, Mark and Siraichi, Marcos Yukio and Suk, Helen and Suo, Michael and Tillet, Phil and Wang, Eikan and Wang, Xiaodong and Wen, William and Zhang, Shunting and Zhao, Xu and Zhou, Keren and Zou, Richard and Mathews, Ajit and Chanan, Gregory and Wu, Peng and Chintala, Soumith},
booktitle = {29th ACM International Conference on Architectural Support for Programming Languages and Operating Systems, Volume 2 (ASPLOS '24)},
doi = {10.1145/3620665.3640366},
month = apr,
publisher = {ACM},
title = {{PyTorch 2: Faster Machine Learning Through Dynamic Python Bytecode Transformation and Graph Compilation}},
year = {2024}
}

@misc{Tensorflow2015,
title={ {TensorFlow}: Large-Scale Machine Learning on Heterogeneous Systems},
url={https://www.tensorflow.org/},
note={Software available from tensorflow.org},
author={
    Mart\'{i}n~Abadi and
    Ashish~Agarwal and
    Paul~Barham and
    Eugene~Brevdo and
    Zhifeng~Chen and
    Craig~Citro and
    Greg~S.~Corrado and
    Andy~Davis and
    Jeffrey~Dean and
    Matthieu~Devin and
    Sanjay~Ghemawat and
    Ian~Goodfellow and
    Andrew~Harp and
    Geoffrey~Irving and
    Michael~Isard and
    Yangqing Jia and
    Rafal~Jozefowicz and
    Lukasz~Kaiser and
    Manjunath~Kudlur and
    Josh~Levenberg and
    Dandelion~Man\'{e} and
    Rajat~Monga and
    Sherry~Moore and
    Derek~Murray and
    Chris~Olah and
    Mike~Schuster and
    Jonathon~Shlens and
    Benoit~Steiner and
    Ilya~Sutskever and
    Kunal~Talwar and
    Paul~Tucker and
    Vincent~Vanhoucke and
    Vijay~Vasudevan and
    Fernanda~Vi\'{e}gas and
    Oriol~Vinyals and
    Pete~Warden and
    Martin~Wattenberg and
    Martin~Wicke and
    Yuan~Yu and
    Xiaoqiang~Zheng},
  year={2015},
}

@article{Numpy2020,
 title         = {Array programming with {NumPy}},
 author        = {Charles R. Harris and K. Jarrod Millman and St{\'{e}}fan J.
                 van der Walt and Ralf Gommers and Pauli Virtanen and David
                 Cournapeau and Eric Wieser and Julian Taylor and Sebastian
                 Berg and Nathaniel J. Smith and Robert Kern and Matti Picus
                 and Stephan Hoyer and Marten H. van Kerkwijk and Matthew
                 Brett and Allan Haldane and Jaime Fern{\'{a}}ndez del
                 R{\'{i}}o and Mark Wiebe and Pearu Peterson and Pierre
                 G{\'{e}}rard-Marchant and Kevin Sheppard and Tyler Reddy and
                 Warren Weckesser and Hameer Abbasi and Christoph Gohlke and
                 Travis E. Oliphant},
 year          = {2020},
 month         = sep,
 journal       = {Nature},
 volume        = {585},
 number        = {7825},
 pages         = {357--362},
 doi           = {10.1038/s41586-020-2649-2},
 publisher     = {Springer Science and Business Media {LLC}}
}

@inproceedings{Cupy2017,
  author       = "Okuta, Ryosuke and Unno, Yuya and Nishino, Daisuke and Hido, Shohei and Loomis, Crissman",
  title        = "CuPy: A NumPy-Compatible Library for NVIDIA GPU Calculations",
  booktitle    = "Proceedings of Workshop on Machine Learning Systems (LearningSys) in The Thirty-first Annual Conference on Neural Information Processing Systems (NIPS)",
  year         = "2017",
  url          = "http://learningsys.org/nips17/assets/papers/paper_16.pdf"
}

@software{Jax2018,
  author = {James Bradbury and Roy Frostig and Peter Hawkins and Matthew James Johnson and Chris Leary and Dougal Maclaurin and George Necula and Adam Paszke and Jake Vander{P}las and Skye Wanderman-{M}ilne and Qiao Zhang},
  title = {{JAX}: composable transformations of {P}ython+{N}um{P}y programs},
  url = {http://github.com/google/jax},
  version = {0.3.13},
  year = {2018},
}

@misc{nvidiaGEMM,
  title={ {User's Guide | NVIDIA Docs: Matrix Multiplication Background}},
  url={https://docs.nvidia.com/deeplearning/performance/pdf/Matrix-Multiplication-Background-User-Guide.pdf},
  author={NVIDIAUsersGuide},
  year={2023},
}

@article{Lewis2022,
  author = {Adam G. M. Lewis  and Jackson Beall  and Martin Ganahl  and Markus Hauru  and Shrestha Basu Mallick  and Guifre Vidal },
  title = {Large-scale distributed linear algebra with tensor processing units},
  journal = {Proceedings of the National Academy of Sciences},
  volume = {119},
  number = {33},
  pages = {e2122762119},
  year = {2022},
  doi = {10.1073/pnas.2122762119}
}

\end{document}